\documentclass{aastex61}
\usepackage{graphicx,natbib}
\usepackage[percent]{overpic}
\usepackage{graphicx,subfigure,amsmath}
\usepackage{hyperref}
\newcommand\gsim{\,\lower3pt\hbox{$\sim$}\llap{\raise2pt\hbox{$>$}}\,}
\newcommand\lsim{\,\lower3pt\hbox{$\sim$}\llap{\raise2pt\hbox{$<$}}\,}

\shorttitle{eruption of flux ropes}
\shortauthors{Fan}

\begin{document}

\title{Simulations of prominence eruption preceded with large amplitude longitudinal oscillations and draining}

\correspondingauthor{Yuhong Fan}
\email{yfan@ucar.edu}

\author{Yuhong Fan}
\affil{High Altitude Observatory, National Center for Atmospheric Research, \\
3080 Center Green Drive, Boulder, CO 80301, USA}

\begin{abstract}
We present magnetohydrodynamic (MHD) simulations of the evolution from quasi-equilibrium to eruption of a prominence-forming twisted coronal flux rope under a coronal streamer. We have compared the cases with and without the formation of prominence condensations, and the case where prominence condensations form but we artificially initiate the draining of the prominence. We find that the prominence weight has a significant effect on the stability of the flux rope, and can significantly increase the loss-of-equilibrium height. The flux rope can be made to erupt earlier by initiating draining of the prominence mass. We have also performed a simulation where large amplitude longitudinal oscillations of the prominence are excited during the quasi-static phase. We find that the gravity force along the magnetic field lines is the major restoring force for the oscillations, in accordance with the ``pendulum model'', although the oscillation periods are higher (by about 10\% to 40\%) than estimated from the model because of the dynamic deformation of the field line dips during the oscillations. The oscillation period is also found to be slightly smaller for the lower part of the prominence in the deeper dips compared to the upper part in the shallower dips. The oscillations are quickly damped out after about 2-3 periods and are followed by prominence draining and the eventual eruption of the prominence.  However we do not find a significant enhancement of the prominence draining and earlier onset of eruption with the excitation of the prominence oscillations compared to the case without.
\end{abstract}

\keywords{magnetohydrodynamics(MHD) --- methods: numerical --- Sun: corona --- Sun: coronal mass ejections (CMEs) --- Sun: filaments, prominences}

\section{Introduction}
Both observations and theoretical analysis have suggested that the weight of
the prominence/filament mass can have a significant role in the stability and
eruption of the hosting magnetic structure that supports it
\citep[e.g.][]{Low:1996, Seaton:etal:2011, Jenkins:etal:2017, Jenkins:etal:2019}.
Using 3D reconstruction of an eruption on 2010 April 3 based on
observations from multiple viewpoints, \citet{Seaton:etal:2011} showed that
mass off-load in an underlying filament triggers the slow rise of a coronal
flux rope, which then reaches a critical height for a catastrophic loss of
equilibrium and produces a coronal mass ejection (CME).
\citet{Jenkins:etal:2019} studied the effect of prominence mass and
mass draining using an analytical model of a flux rope as a line
current suspended within a background potential field.
They showed that the inclusion of the prominence mass can increase the
height at which the line current experiences an ideal-MHD instability
or loss of equilibrium. With mass draining, which allows the flux rope
to rise, the critical height for the loss of equilibrium can occur at
a range of heights depending explicitly on the amount and evolution
of the mass.

Besides potentially playing an important role in the development
of eruption, prominence plasma exhibits various dynamic phenomena
which can provide information about the conditions of the hosting
coronal magnetic structure, which is difficult to observe directly. 
One interesting dynamic phenomenon is the large amplitude longitudinal
(LAL) oscillations first reported by \citet{Jing:etal:2003}.
These are coherent oscillations of the prominence
material (nearly) along the filament axis with amplitudes of
$30$ - $100 \, {\rm km/s}$ and periods of $0.83$ - $2.66 \, {\rm hours}$,
triggered by an energetic event close to the filament footpoints
\citep[e.g.][]{Luna:Karpen:2012,Tripathi:etal:2009}.
\citet{Luna:Karpen:2012} developed the ``pendulum model'' to explain
the LAL oscillations, where the projected gravity force along the magnetic field
lines is the restoring force for the prominence condensations to oscillate
in the magnetic concavity of the field line dips, and the period of the
oscillations is determined by the curvature radius of the dips,
analogous to the length of the pendulum.
These oscillations therefore provide a useful means of diagnosing the
conditions of the magnetic support of the prominence and can put
constraint on the magnetic field strength \citep{Luna:Karpen:2012}.
The LAL oscillations are not necessarily associated with a subsequent
eruption \citep[e.g.][]{Jing:etal:2003, Jing:etal:2006, Vrsnak:etal:2007}.
However, \citet{Bi:etal:2014} reported the observation of a filament
eruption where both LAL oscillations followed by filament material draining
are found to precede the eruption.  This raises the question of the role LAL
oscillations may play for the initiation of the draining and the eruption. 

Increasingly, 3D MHD simulations of CMEs and CME source regions are
conducted with more realistic treatment of the thermodynamics to allow
for the formation of the prominence condensations to study their dynamic
effects and observable signatures
\citep[e.g.][]{Xia:etal:2014,Xia:Keppens:2016a,Fan:2017,Fan:2018,
Zhou:etal:2018}.
\citet{Xia:etal:2014} and \citet{Xia:Keppens:2016a} have carried out
the first 3D MHD simulations of the formation of a prominence in a stable
equilibrium coronal flux rope, with the inclusion of the non-adiabatic
effects of an empirical coronal heating, optically thin radiative losses,
and field-aligned thermal conduction. They have used an adaptive
grid to resolve the fine-scale internal dynamics of the prominence, and
reproduced many observed features seen in SDO/AIA observations of
the prominence-cavity systems.
\citet{Zhou:etal:2018} have performed 3D MHD simulations of prominence
oscillations (both longitudinal and transverse oscillations) in a stable
magnetic flux rope, and compared the results with idealized analytical
models. They excite the oscillations by imparting initial velocity
perturbations to the prominence.  They found that the resulting LAL
oscillations are in agreement with the pendulum model where the field-aligned
component of the gravity serves as the restoring force. The period of
the LAL oscillations are higher than that predicted by the pendulum model by up to 20\%.
\citet{Fan:2017, Fan:2018} have carried out 3D MHD simulations of
a prominence-forming coronal flux rope that
transitions from quasi-equilibrium to eruption.
The thermodynamics treatment incorporates the
non-adiabatic effects of a simple empirical coronal heating, optically thin
radiative losses, and field-aligned thermal conduction.
In these simulations a significantly twisted, long coronal flux rope
builds up under a pre-existing coronal streamer by an imposed flux emergence
at the lower boundary.
Cool prominence condensations form in the dips of the long emerged twisted
field lines due to in-situ radiative instability driven by the optically
thin radiative cooling.  The prominence-carrying flux rope undergoes a long
quasi-static rise phase, and develops prominence draining during the later
stage as the dips become shallower with the rise \citet{Fan:2018}. The
flux rope eventually erupts and develops an associated prominence eruption
when the center portion of the flux rope rises to a certain height.
It is found that once the prominence is formed, the magnetic
field supporting the prominence becomes significantly non-force-free,
despite the fact that the entire flux rope has low plasma-$\beta$. 
Through a comparison of the simulations with and without the formation of the
prominence condensations (the ``PROM'' and ``non-PROM'' simulations in
\citet{Fan:2018}), it is found that the prominence weight can
suppress the development of the kink instability of the highly twisted
emerged flux rope and delay its rise to the critical height for the loss
of equilibrium and dynamic eruption of the flux rope \citep{Fan:2018}.

In this paper we expand upon the study of \citep{Fan:2018},
and carry out further simulations to investigate the dynamic effects of
prominence draining and prominence LAL oscillations.
Further comparison of the simulations with and without prominence formation
shows that the presence of the prominence weight also causes a significant
increase of the loss-of-equilibrium height of the flux rope, consistent with
the result from the analytical model by \citet{Jenkins:etal:2019}.
A new simulation where we artificially initiate prominence draining by reducing
the pressure at one footpoint of the flux rope during
the quasi-static rise phase, shows that a significant reduction of the total
prominence mass allows the flux rope to rise more quickly to the loss-of-equilibrium
height and develop a dynamic eruption significantly earlier.
We also perform a new simulation where we excite LAL oscillations by adding
an initial velocity (parallel to the magnetic field) to the prominence
plasma during the quasi-static rise phase.
We find prominence LAL oscillations with a period of roughly 2 hours develop
in the magnetic dips. It is found that the main restoring force of the oscillations
is the field-aligned gravity force consistent with the pendulum model, although the
oscillation period is found to be larger than predicted by the model.  
The oscillations are strongly damped and die out after about 2-3 periods.
They are followed by episodes of prominence draining towards the two ends of the flux
rope and an eventual dynamic eruption of the prominence. However the excitation
of the LAL oscillations is not found to significantly enhance the draining and
cause an earlier eruption compared to the case without the LAL oscillations.

\section{Model Description}
\label{sec:model}

The MHD numerical simulations we use in this work are specifically the ``PROM''
and ``non-PROM'' simulations described in \citet[][hereafter F18]{Fan:2018},
and two additional simulations: ``PROM-drain'' with induced prominence draining
and ``PROM-LALO'' with induced LAL oscillations, by
modifications to the previous ``PROM'' simulation as described below.
The readers are referred to section 2 in F18
as well as sections 2 and 3.1 in \citet[][hereafter F17]{Fan:2017}
for a detailed description of the setup of the previous ``PROM'' and
``non-PROM'' simulations and the numerical model.
As a summary, we use the the
``Magnetic Flux Eruption'' (MFE) code to solve the set of
semi-relativistic MHD equations in spherical geometry (F17).
The energy equation explicitly incorporates the non-adiabatic effects
of a simple empirical coronal heating (that depends on height only),
optically thin radiative cooling, and the field-aligned heat conduction.
The simulation domain is in the corona, excluding the photosphere and
chromosphere layers, with the lower boundary temperature and density
set at the transition region.
With the inclusion of the above non-adiabatic effects, the simulations
allow in-situ formation of prominence condensations in the
the corona as a result of the development of the radiative
instability.
The simulations are carried out in a spherical wedge domain
with a radial range $r \in [R_s, 11.47 R_s]$,
a colatitude range $\theta \in [75^{\circ}, 105^{\circ}]$,
and an azimuthal range $\phi \in [-75^{\circ}, 75^{\circ}]$,
where $R_s$ is the solar radius.
The domain is first initialized with a 2D quasi-steady
solution of a coronal steamer with an ambient solar wind
(section 3.1 in F18).
At the lower boundary, the emergence of a portion of a twisted magnetic
torus is imposed as described in F17 (equations (19)-(22) in the paper)
such that a long twisted flux rope is built up quasi-statically under
the streamer dome. The same flux emergence is imposed in the ``PROM'' and
``non-PROM'' simulations in F18.  An extended prominence
condensation develops in the dips of the emerged flux rope field lines
in the ``PROM'' case, whereas in the
``non-PROM'' case, the prominence formation is suppressed by modifying
the radiative cooling and thermal conduction (F18).  The resulting
subsequent evolution of the flux rope is drastically different between
the two cases as described in F18.

Here we further carry out two simulations labeled ``PROM-drain''
and ``PROM-LALO'' cases.
For the ``PROM-drain'' simulation we artificially initiate draining of
the prominence at a time instance during the ``PROM'' simulation after
the prominence has formed in the emerged flux rope but while it is still
in the quasi-static phase of the evolution.  We do this by modifying
the pressure at the lower boundary at one footpoint of the flux rope.
As described in F17, we adjust the base pressure at the lower boundary
(as given by eqs. [17] and [18] in F17) such that it is driven towards
a value that is proportional to the downward heat conductive flux to
crudely represent the effect of chromospheric evaporation. To facilitate
the draining, we reduce the constant of proportionality $C$ for the
base pressure in equation (18) in F17 by about 50 times for a period of
about 9.9 hours (from about $t=17.2$ hour to $t=27.1$ hour) during the
quasi-static evolution phase, over a region that encloses the right
foot-point of the emerged flux rope, in the original ``PROM'' simulation.
For the ``PROM-LALO'' simulation, we initiate LAL oscillations
by imparting an initial parallel momentum (parallel to the direction of
the magnetic field lines) to the cool prominence mass at a time instance
during the quasi-static phase in the original ``PROM'' simulation.
At about $t=17.15$ hour, we add a velocity parallel to the magnetic field
of about $100$ km/s to the cool plasma where temperature $T < 10^5$ K.
We examine the subsequent evolutions of the ``PROM-drain'' and ``PROM-LALO''
simulations after their respective initiation of the perturbations and
compare them with the evolution of the original ``PROM'' simulation.

\section{Simulation Results}
\label{sec:results}

\subsection{The effect of prominence weight and initiation of eruption by prominence draining}
\begin{figure}[htb!]
\centering
\includegraphics[width=0.75\textwidth]{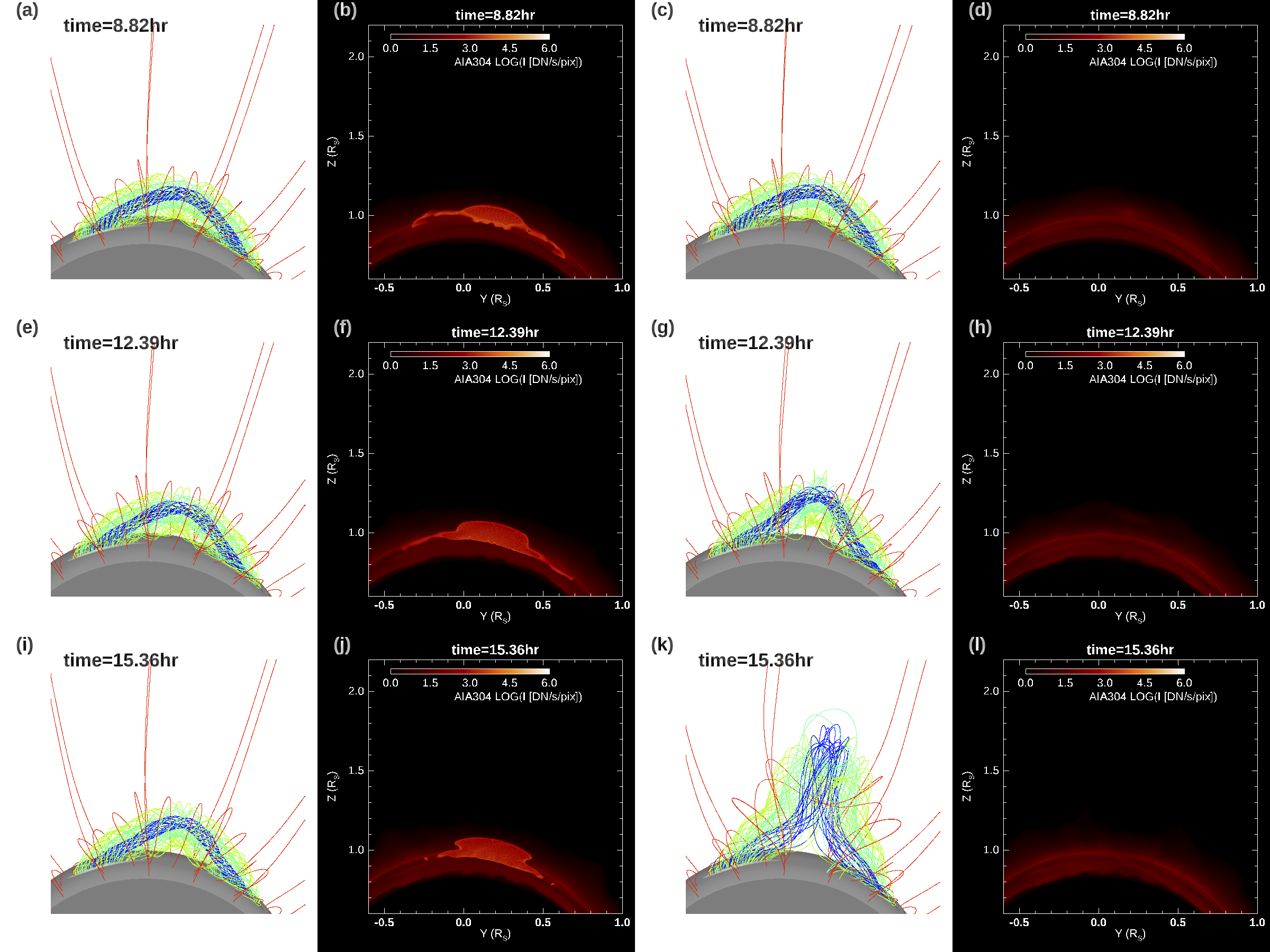}
\caption{Side-by-side comparison of the evolution from the ``PROM'' and ``non-PROM'' simulations.
The 1st and 2nd columns show snapshots of the 3D magnetic field lines
and the corresponding synthetic SDO/AIA 304 {\AA} emission images from the ``PROM''
simulation, and the 3rd and 4th columns show those at the concurrent times for the
``non-PROM'' simulation.
An \href{https://drive.google.com/file/d/13I20b03b8gjoEes89tB_yHF9gUoiad5W/view?usp=sharing}{\bf animation}
of the evolution for the two cases in comparison is also
available in the online version of the paper. The animation shows the
evolution of both models from $t=5.95$ to $22.30$ hour, and continues
the PROM simulation from $22.30$ to $39.64$ hour.}
\label{fig:prom_nonprom_fdl_304}
\end{figure}

First, as an overview of the result from F18 on the ``PROM'' and ``non-PROM'' simulations,
Figure \ref{fig:prom_nonprom_fdl_304}
(and the associated animation in the online version) shows a side-by-side comparison of the
evolutions produced by the ``PROM'' simulation with the formation of the prominence
condensations (1st and 2nd columns) and the ``non-PROM'' simulation without prominence
formation (3rd and 4th columns).
The 2nd and 4th columns show the synthetic SDO/AIA 304 {\AA} images
corresponding to the 3D field-line images to their left (with the same perspective view).  
The synthetic SDO/AIA 304 {\AA} channel emission images are computed by line-of-sight (LOS)
integrations through the simulation domain using equation (23) in F17, and they show
concentrations of cool plasma with a peak temperature response at about $8 \times 10^4$ K.
In carrying out the LOS integrations here, we have also assumed that the prominence
condensations are “opaque” such that when the LOS reaches a plasma where both the
temperature goes below $7.5 \times 10^4$ K and the number density is above $10^9 {\rm cm}^{-3}$,
we stop the integration for that LOS assuming the emission from behind the
plasma is blocked and does not contribute to the integrated emission for the LOS.
As described in F18, for both the ``PROM'' and ``non-PROM'' cases, the emergence of
an identical twisted flux rope is imposed at the lower boundary and the emergence
is stopped (at $t=8.82$ hr, panels (a),(b),(c),(d)), when the total field line twist about
the axis of the emerged flux rope reaches 1.83 winds, which is above the critical
twist (1.25 winds) for the onset of the kink instability for a line-tied, uniformly
twisted cylindrical force-free flux tube \citep{Hood:Priest:1981}.
In the ``PROM'' case, a long prominence has formed in the field line dips of the
emerged flux rope (panels (a) (b)), whereas no prominence forms in the ``non-PROM'' case
(panels (c) (d)), and the subsequent evolution is found to be very different for
the two cases.  In the ``non-PROM'' case, the flux rope quickly develops the kink motion
with the central portion protruding upward and then erupts dynamically when it reaches
a certain height (panels (g) (k) and the online movie).
In contrast during this same period, the prominence carrying
flux rope in the ``PROM'' case remains confined, showing very little motion (panels (e) (f)
and the online movie).  It is found that the flux rope undergoes a long quasi-static rise
phase (of about 30 hours) with episodes of prominence draining before eventually its
center portion rises to a certain height where it erupts dynamically, with an associated
prominence eruption and draining along the two legs of the erupting flux rope (see the
online movie and also F18).  F18 found that the prominence carrying magnetic field in
the flux rope is significantly non-force-free despite the low plasma $\beta$.  The weight
of the prominence suppresses the development of the kink instability and delays the flux
rope's rise to the critical height for the loss of equilibrium.

\begin{figure}[htb!]
\centering
\includegraphics[width=0.75\textwidth]{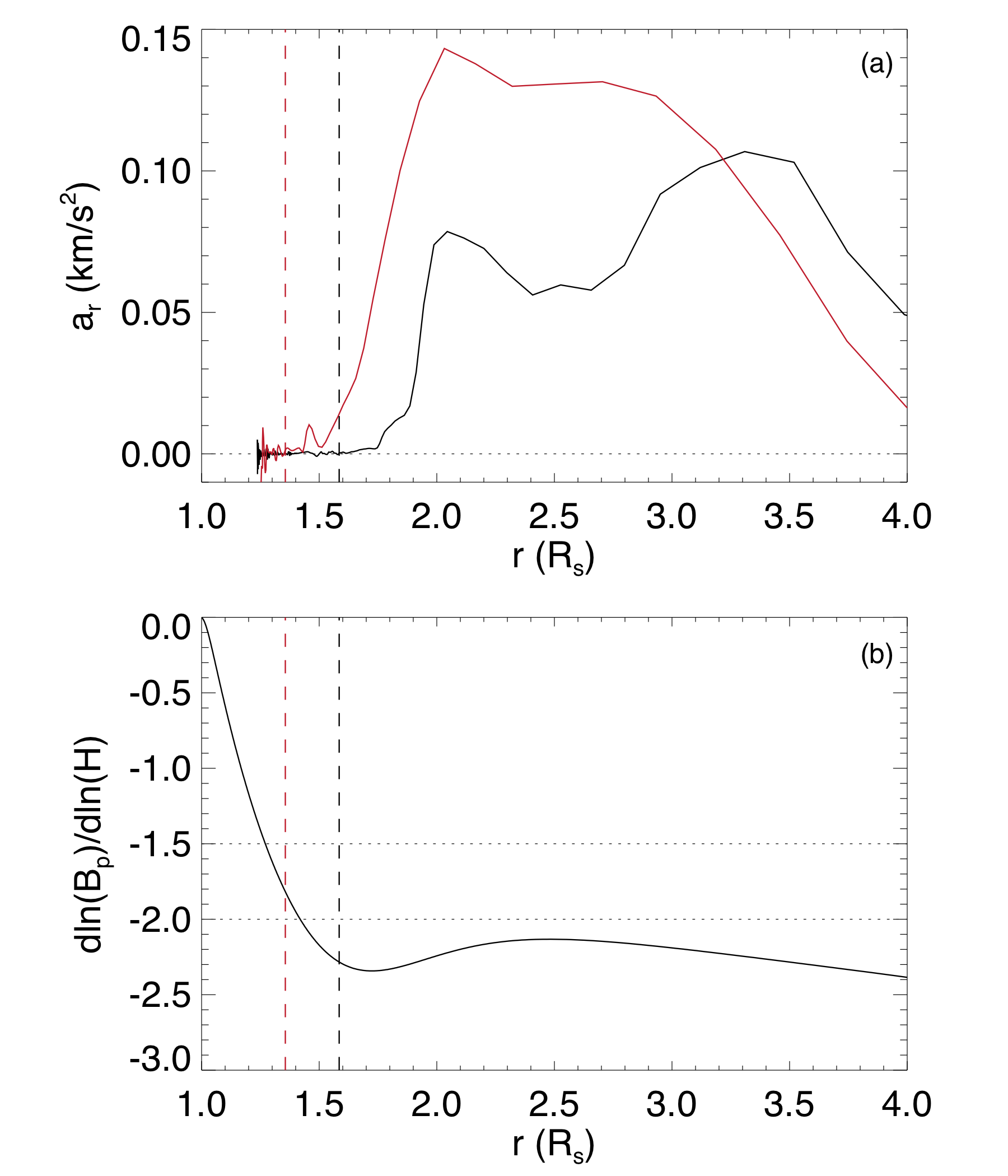}
\caption{(a) Acceleration as a function of height tracked at the apex of
the axial field line of the flux rope for the ``PROM'' case (black curve)
and the ``non-PROM'' case (red curve); (b) The decay rate with height of
the corresponding potential magnetic field after the flux emergence is stopped.
The vertical dashed lines mark the height at which the acceleration becomes
persistently positive, approximating the loss-of-equilibrium height, for
the ``PROM'' case (black dashed line) and the ``non-PROM'' case (red dashed line).}
\label{fig:bpdecay_ar_compare}
\end{figure}

In this paper, we further show that the loss-of-equilibrium height of the flux rope is
significantly increased for the ``PROM'' case compared to the ``non-PROM'' case.
Figure \ref{fig:bpdecay_ar_compare} shows the acceleration as a function of height
tracked at the apex of the axial field line of the flux rope.
We find that the height at which the acceleration becomes persistently positive,
which approximates the loss-of-equilibrium height,
is significantly higher for the ``PROM'' case than that for the ``non-PROM'' case.
The decay rate with height of the corresponding potential field at the height for
the loss of equilibrium for the ``PROM'' case (see Figure \ref{fig:bpdecay_ar_compare}
at the location marked by the black dashed line) is about 2.3, which is also
significantly higher than the critical value (of about 1.5) for the onset of the torus
instability for a toroidal current ring \citep[e.g.][]{Kliem:Toeroek:2006,
Demoulin:Aulanier:2010}. An analytic study by \citet{Jenkins:etal:2019}, which models
a flux rope as a line current confined in a background potential magnetic field,
showed that the inclusion of the prominence weight increases the height at which
the line current experiences loss of equilibrium.  The basic reason is that the
additional (constant with height) weight of the prominence causes the total confining
force to decline more slowly with height, such that the flux rope needs to reach a
higher height where the potential field has a steeper decline with height (than that
is needed in the absence of the prominence weight) for the loss of equilibrium or
torus instability to take place. Our simulation result of the increased 
loss of equilibrium height for the flux rope with the formation of prominence
is consistent with the result of the analytic model by \citet{Jenkins:etal:2019}.

\begin{figure}[htb!]
\centering
\includegraphics[width=0.75\textwidth]{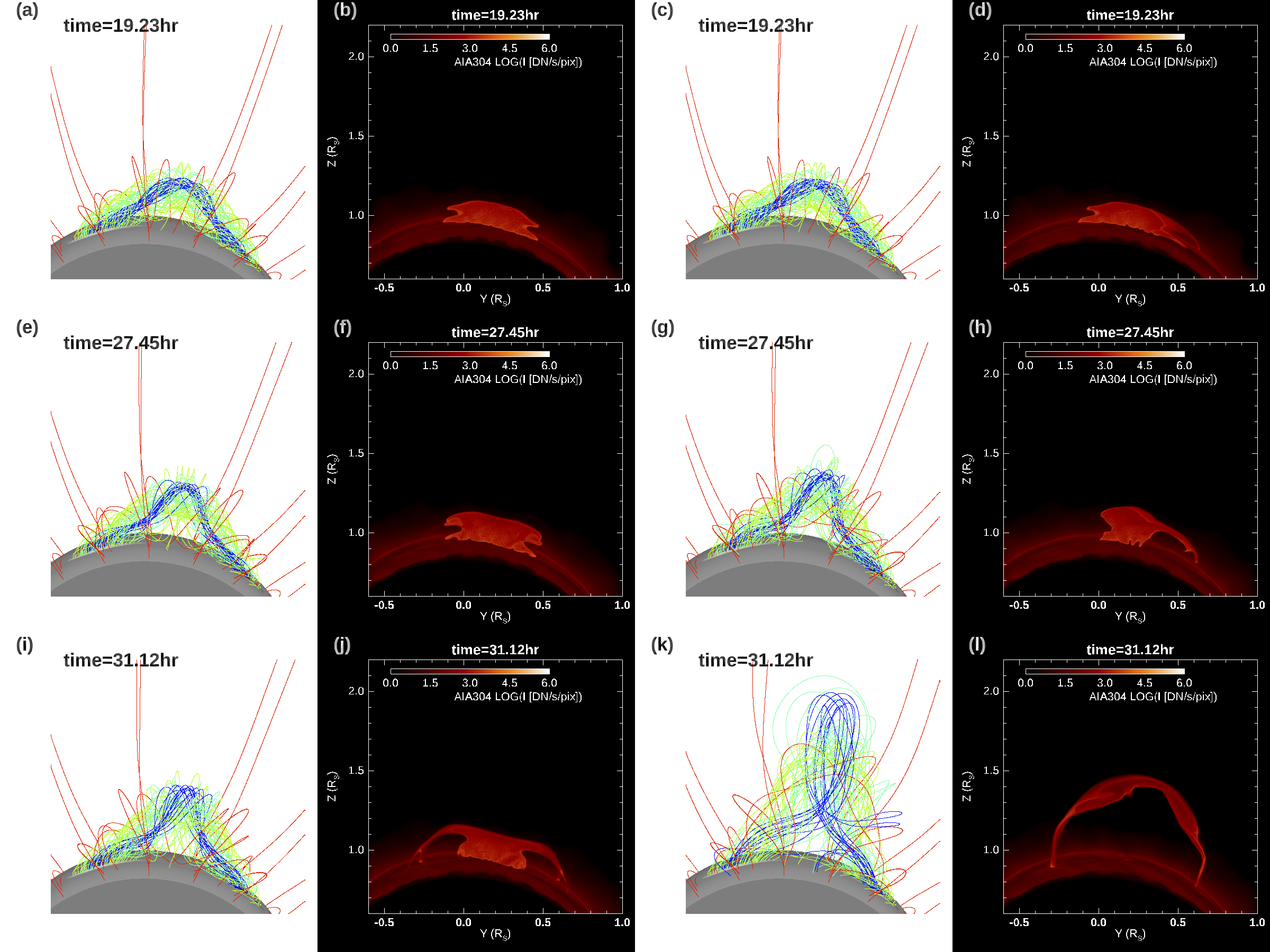}
\caption{Comparison of the evolution from the ``PROM'' and ``PROM-drain'' simulations.
The left two columns show snapshots of the 3D magnetic field lines
and the corresponding synthetic SDO/AIA 304 {\AA} emission images from the ``PROM''
simulation, and the right two columns show those at the concurrent times for the
``PROM-drain'' simulations.
An \href{https://drive.google.com/file/d/1k8R3ArSVdvbzJD8-wbprbQnB80DWCpny/view?usp=sharing}{\bf animation}
of the evolution for the two cases in comparison is also available in the online version of the paper.
The animation shows the evolution of both models from $t=17.15$ to $33.90$ hour,
and continues the PROM simulation from $33.90$ to $39.64$ hour.}
\label{fig:prom_promdrain_fdl_304}
\end{figure}

\begin{figure}[htb!]
\centering
\includegraphics[width=0.5\textwidth]{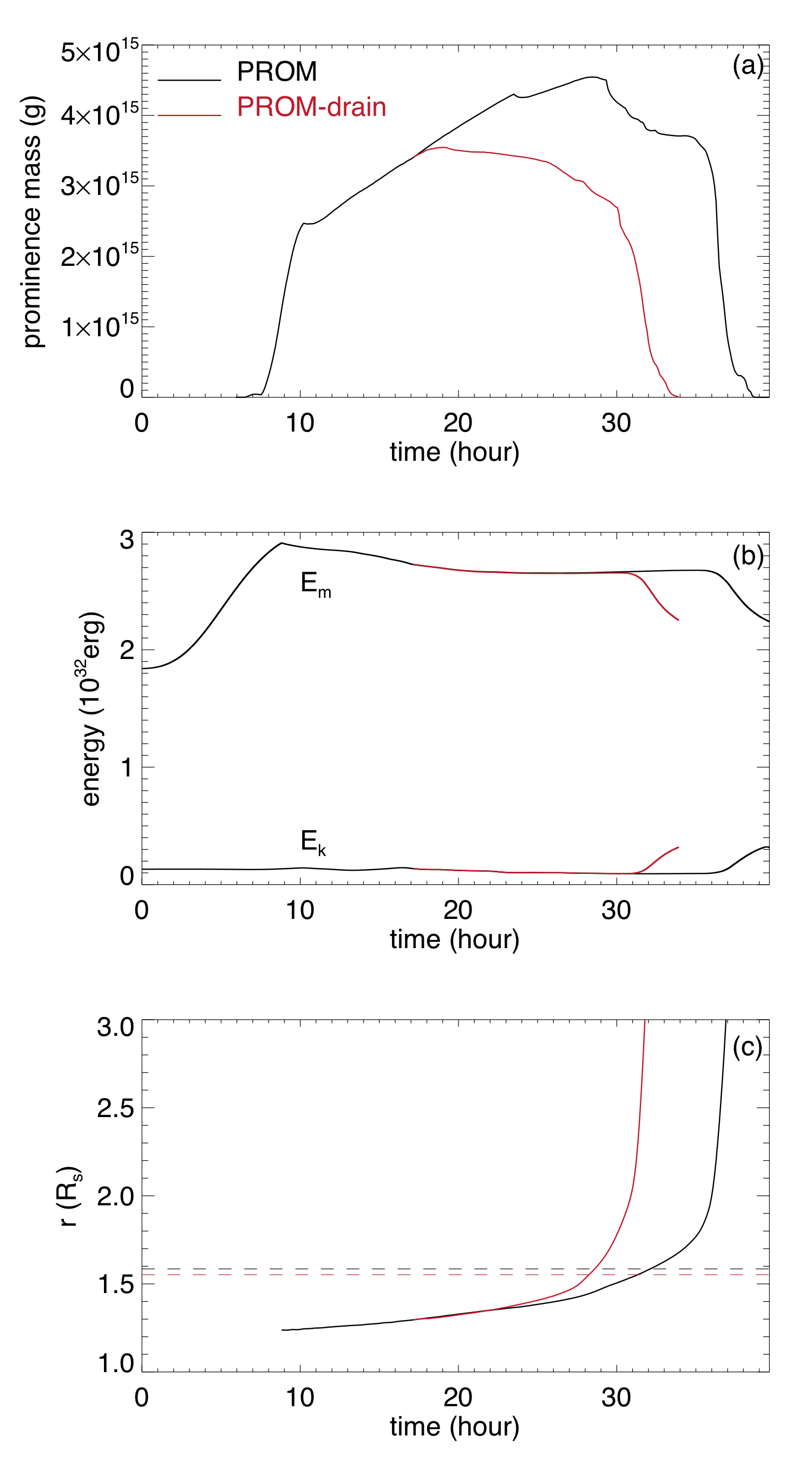}
\caption{(a) The temporal evolution of the cool prominence mass in the corona evaluated as the
total mass with temperature below $10^5$ K; (b) The temporal evolution of the total magnetic
energy $E_m$ and total kinetic energy $E_k$; (c) Height vs. time tracked at the apex of the
axial field line of the emerged flux rope. In each panel, the red (black) solid curve
shows the result for the ``PROM-drain'' (``PROM'') case. The red (black) dashed line in (c)
marks the loss-of-equilibrium height for the ``PROM-drain'' (``PROM'') case, which is
approximated by the height at which persistently positive acceleration begins.}
\label{fig:prommass_rvst_evol}
\end{figure}

To further investigate the effect of prominence draining, we have carried out the
``PROM-drain'' simulation where we artificially initiate prominence draining 
at a time ($t=17.15$) during the quasi-static rise phase of the ``PROM'' simulation
 by reducing the pressure at the right foot point of the flux rope, as described
in section \ref{sec:model}.  Figure \ref{fig:prom_promdrain_fdl_304} shows
the subsequent evolution of the ``PROM-drain'' case (3rd and 4th columns) compared
the ``PROM'' case (1st and 2nd columns). An animation of the this side by side
comparison of the evolution is also available in the online version.
As can be seen from the Figure (and the movie) that an earlier draining
of prominence mass towards
the right footpoint (Figure \ref{fig:prom_promdrain_fdl_304}(d)) takes place due to the
lowered pressure at the right footpoint of the flux rope while no draining has yet
happened in the ``PROM'' case.  As a result the quasi-static rise is faster in
the ``PROM-drain'' case and it develops an earlier eruption
(Figures \ref{fig:prom_promdrain_fdl_304}(g)(h)(k)(i)).
On the other hand in the ``PROM'' case, prominence draining also takes place but
significantly later, and the flux rope develops a later eruption with
an associated prominence eruption (see the movie associated with Figure \ref{fig:prom_promdrain_fdl_304}).
Figure \ref{fig:prommass_rvst_evol}(a) shows the evolution of the prominence mass,
evaluated as the total mass with temperature below $10^5$ K, comparing the two cases.
We see that starting from the time $t=17.15$ hour (when the pressure at the right foot point
of the flux rope is lowered) the prominence mass for the ``PROM-drain'' case (the red curve)
starts to decline, while the prominence mass for the ``PROM'' case continues to rise
due to continued convergence of mass towards the field line dips (F18). The draining
towards the right footpoint of the flux rope in the ``PROM-drain'' case starts an
earlier decline of the prominence weight, causing a faster rise of the flux
rope as shown by the height-time curves of the tracked apex of the axial field line
of the flux rope (Figure \ref{fig:prommass_rvst_evol}(c)). This allows the flux rope in
the ``PROM-drain'' case to reach the loss-of-equilibrium height earlier and develop
an earlier eruption by about 5 hours, resulting in a rapid acceleration, a significant
magnetic energy release and a significant kinetic energy increase
(Figure \ref{fig:prommass_rvst_evol}(b)).
We find that the loss-of-equilibrium height (marked by the dashed lines in
Figure \ref{fig:prommass_rvst_evol}(c)), which we approximate as the height at which
the acceleration at the tracked apex of the axial field line becomes persistently positive,
is close for the two cases, with the ``PROM-drain'' case being slightly lower.
Thus prominence draining can initiate an earlier eruption of the flux rope by allowing it
to rise to the loss-of-equilibrium height more quickly.

\subsection{Eruption preceded with LAL oscillations and draining of prominence}
\begin{figure}[htb!]
\centering
\includegraphics[width=0.75\textwidth]{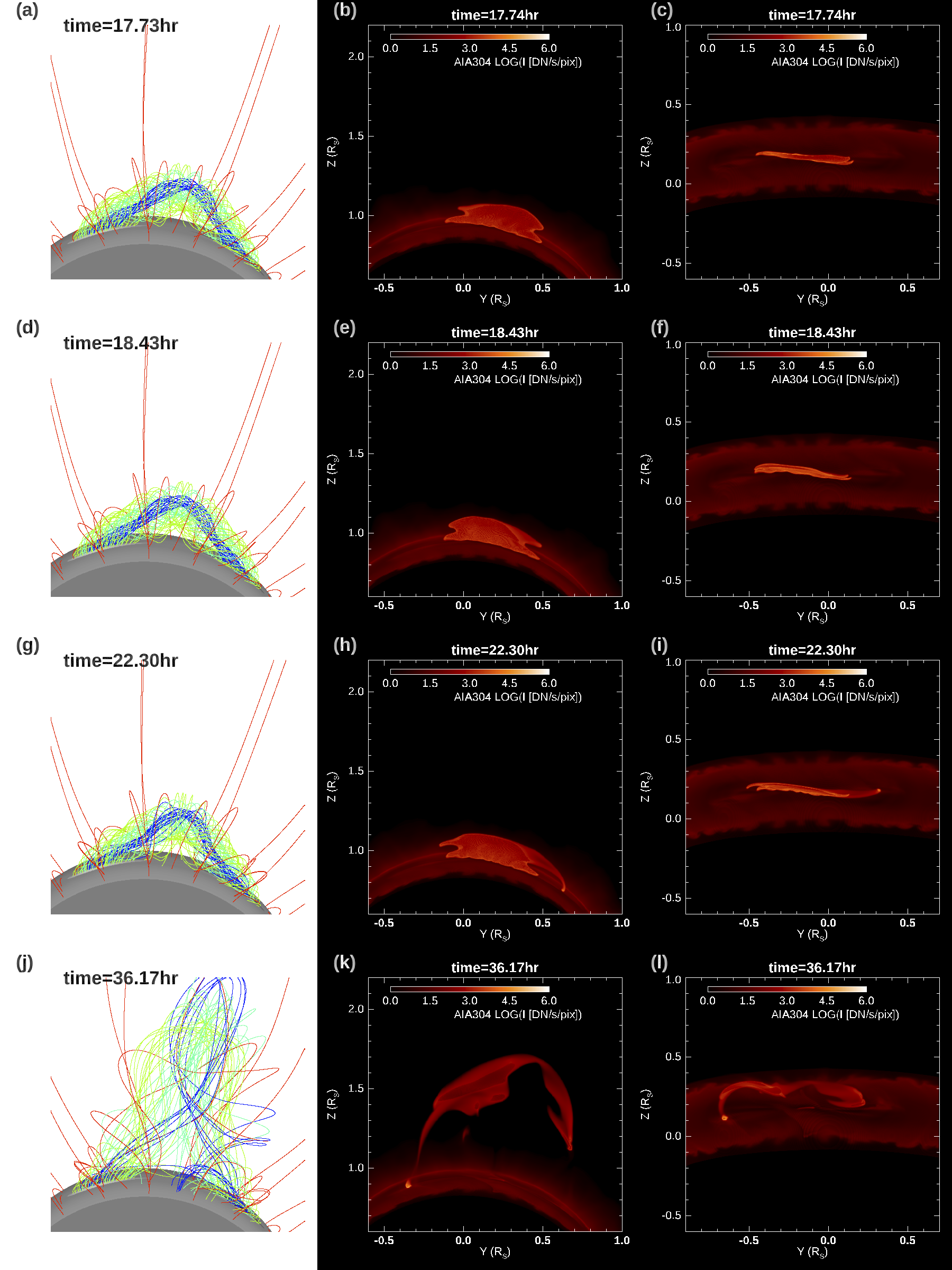}
\caption{Evolution from the ``PROM-LALO'' simulation. The left column shows
snapshots of the 3D magnetic field lines, and the middle and
right columns show the synthetic SDO/AIA 304 {\AA} emission images as viewed
from 2 different perspectives, with the middle column being of the same perspective
as the field line snapshots and the right column for an on-disk view from above. 
An \href{https://drive.google.com/file/d/1ukxtgTibB3lzXy5zCf0habBe3khgRMZ0/view?usp=sharing}{\bf animation}
of the evolution from the time the oscillation is initiated
till the prominence eruption ($t=17.15$ to $39.64$ hour) is available in
the online version of the paper.}
\label{fig:lao_fdl_304}
\end{figure}

Figure \ref{fig:lao_fdl_304} and the associated online movie show the evolution
of the 3D magnetic field and the synthetic SDO/AIA 304 {\AA} emission from
the ``PROM-LALO'' simulation, in which we initiate prominence oscillations by
imparting an initial parallel velocity (parallel to the magnetic field lines)
to the cool prominence mass at a time instance during the quasi-static phase in
the original ``PROM'' simulation.
We see that with the initial velocity imparted to the prominence plasma at
$t=17.15$ hour, the prominence as a whole develops large amplitude
oscillations (panels (b),(e),(c),(f)) with a period of roughly 2 hours,
with the magnetic field also showing horizontal swinging oscillations of the
entire flux rope (see the movie associated with Figure \ref{fig:lao_fdl_304}).
These large amplitude oscillations are
strongly damped and die out after about 2-3 periods. They are then followed
by episodes of prominence draining towards either footpoints as the flux rope
continues to rise quasi-statically (panels (g),(h),(i) and the movie),
until eventually it erupts with an associated prominence eruption
(panels (j),(k),(l)). As can be best seen in the movie of Figure \ref{fig:lao_fdl_304} (the on-disk view
of the synthetic AIA 304 {\AA} images), the displacements of the prominence
condensations during the oscillations are not exactly aligned with the prominence
spine, but are at a small acute angle from the prominence spine. 
This is because the motion of the prominence mass is mainly along the magnetic
field, and magnetic field lines supporting the prominence at the dips are at a
small acute angle relative to the apparent orientation (spine) of the
prominence concentrations \citep[F17,][]{Zhou:etal:2018,Luna:etal:2018}.
It can also be seen from the movie of Figure \ref{fig:lao_fdl_304} (the middle synthetic AIA 304 {\AA} images)
that the prominence oscillations are showing out-of-phase displacements, with
the upper part of the prominence exhibiting a slightly longer oscillation
period than the lower part.  The amplitude of the oscillations is also greater
for the upper part than the lower part.

\begin{figure}[htb!]
\centering
\includegraphics[width=0.75\textwidth]{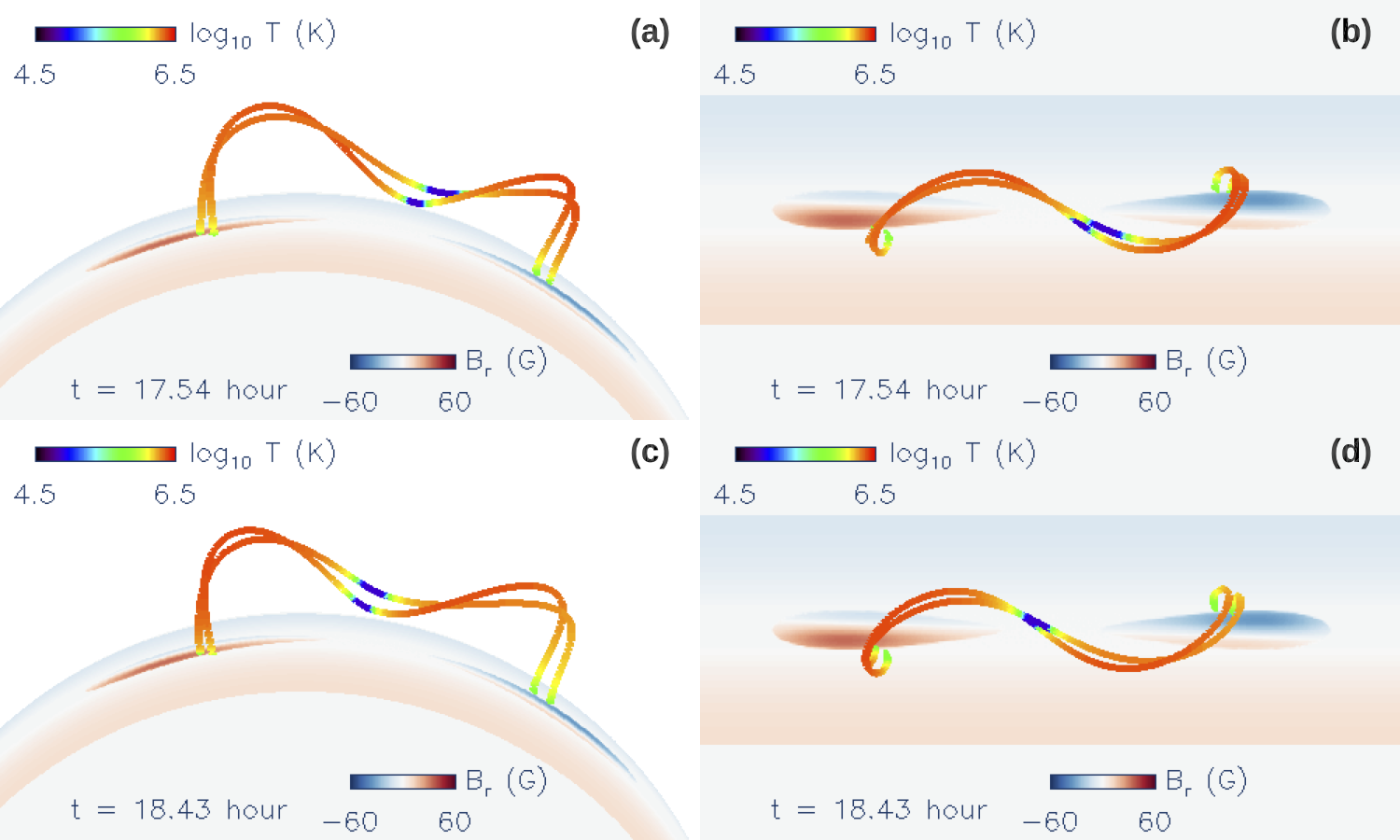}
\caption{Snapshots of two tracked prominence carrying field lines (traced
from two fixed left footpoints) colored in temperature, showing the
oscillation motions of the prominence condensations (mainly) along the
field lines. The left and right columns show two different perspective views.
An \href{https://drive.google.com/file/d/1NRZtPkcmNHoALC5tPR8JQ6m-pMqva_U8/view?usp=sharing}{\bf animation}
of the evolution of the two tracked field lines is also 
available in the online Journal. The animated view runs from $t=17.15$ to 
$22.60$ hour.}
\label{fig:2tempfdl_lalo}
\end{figure}

\begin{figure}[htb!]
\centering
\includegraphics[width=0.75\textwidth]{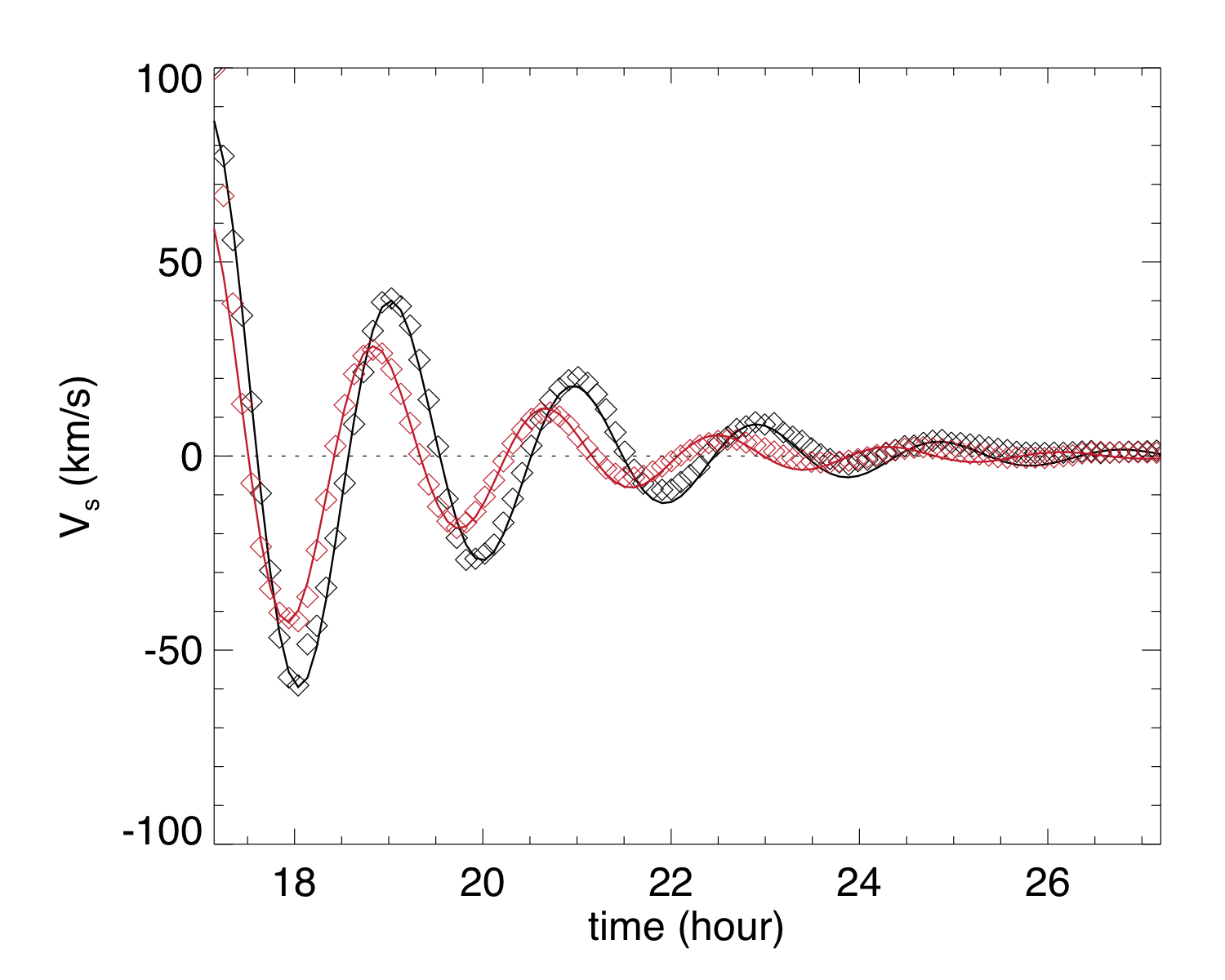}
\caption{Temporal evolution of the parallel velocity
(parallel to the magnetic field) of the prominence
condensation in the higher (black diamond points) and the lower (red diamond
points) dips of the two tracked field lines shown in
Figure \ref{fig:2tempfdl_lalo}. The black and red solid curves are the fitted
exponatially decaying sinusoidal functions to the black and red points.}
\label{fig:laofit}
\end{figure}

\begin{figure}[htb!]
\centering
\includegraphics[width=0.75\textwidth]{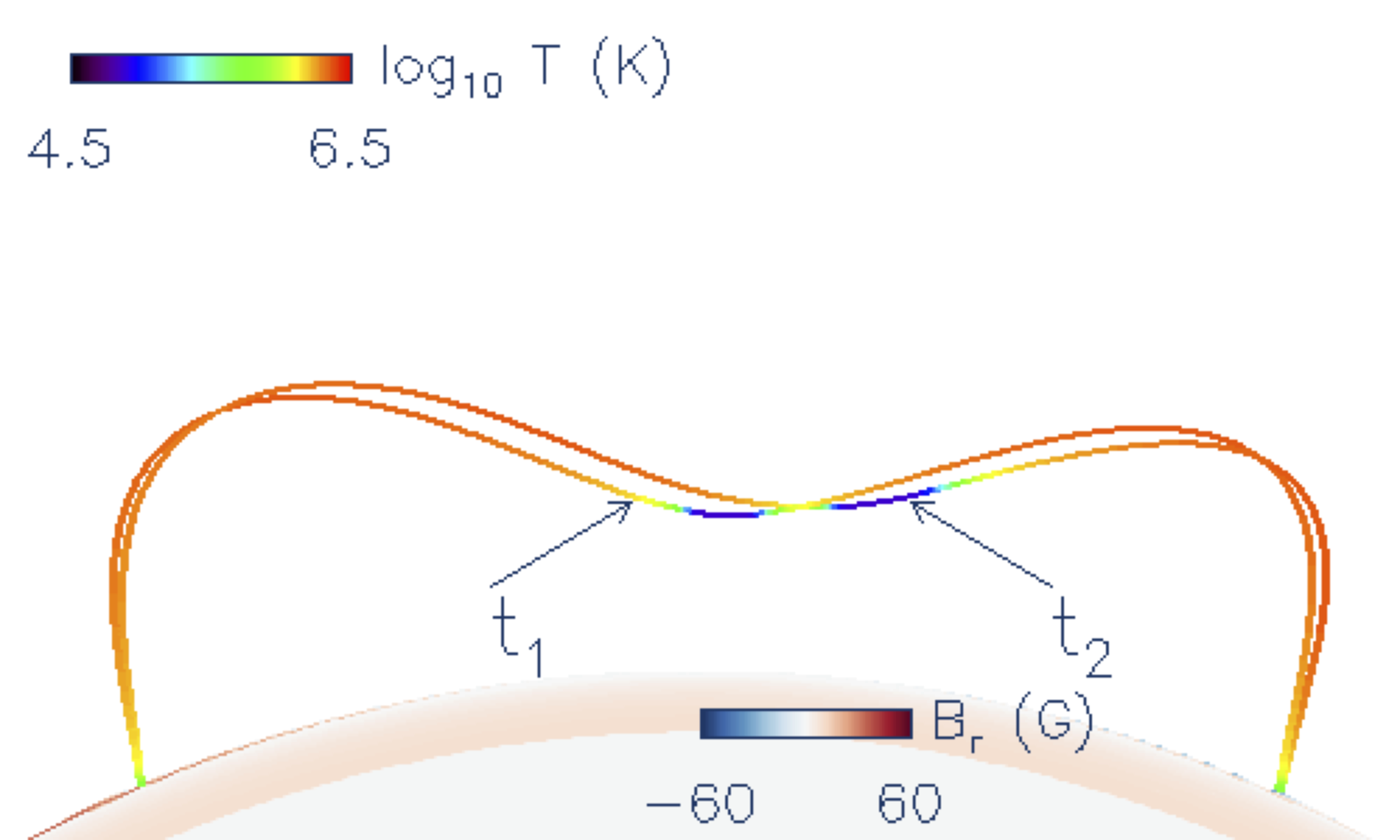}
\caption{A prominence carrying field line at two time instances ($t_1$ and $t_2$)
during the oscillation of the prominence, illustrating the deformation of the
field line.}
\label{fig:2tempfdl_evol_lao}
\end{figure}

Figure \ref{fig:2tempfdl_lalo} and the associated movie show the evolution
of two tracked prominence-carrying field lines colored in temperature, traced
from two fixed left footpoints on the lower boundary.
They show the oscillation motions of the two cool prominence
condensations at the dips of these two field lines. It can be seen that
the motions of the condensations are mainly along the magnetic field lines,
but the field lines are not rigid, with the field line themselves showing
swinging motions.  It can also be seen from the movie that the motions
of the two condensations gradually become out of phase with the condensation
in the higher shallower dip showing a slightly longer oscillation period than
that in the lower deeper dip.
Figure \ref{fig:laofit} shows the parallel velocity (parallel to the magnetic
field) as a function of time of the two prominence condensations shown in Figure 
\ref{fig:2tempfdl_lalo}.
The velocity is measured at the temperature minimum on the tracked field lines.
We have fitted an exponentially decaying sinusoidal function to the measured
velocity:
\begin{equation}
v_s = A_0 \sin \left ( \frac{2 \pi}{P} ( t-t_0 ) - \phi_0 \right )
\exp \left ( - \frac{t-t_0}{\tau} \right ),
\label{eq:fitfunc}
\end{equation}
where $t_0=17.15$ hour is the time the initial velocity is imparted to the prominence
plasma, and $A_0$, $\phi_0$, $P$ and $\tau$ are the fitted initial amplitude, initial
phase, oscillation period, and e-folding decay time respectively.
We obtain oscillation period $P=1.94$ hour ($P=1.82$ hour) and decay time $\tau=2.45$
hour ($\tau=2.20$ hour) for the motion of the higher (lower) dip prominence concentration.
These values are within the observed ranges for the period an decay time for LAL oscillations
of prominences \citep[e.g.][]{Jing:etal:2006,Tripathi:etal:2009,Luna:Karpen:2012,Luna:etal:2018}.
\citet{Luna:Karpen:2012} constructed the ``pendulum model'' to explain the
prominence LAL oscillations, in which the projected gravity along the magnetic field is the
restoring force for the oscillatory motions of the prominence condensations along
the magnetic dips. In this model the oscillation period is given by
\begin{equation}
P_{\rm pendulum}=2 \pi \sqrt{\frac{R_c}{g}},
\label{eq:period}
\end{equation}
where $R_c$ is the radius of curvature of the magnetic
field line dip and $g$ is the gravitational acceleration.
We estimate a mean $R_c$ for the field line dip by averaging the radius
of curvature over a length from the bottom of the dip that corresponds to the
initial peak displacement of the oscillations.  Then the theoretical period based
on the estimated $R_c$ is found to be $P_{\rm pendulum} = 1.76$ hour
($P_{\rm pendulum} = 1.30$ hour) for the higher (lower) dip. The actual
period $P$ found in the simulation above is significantly greater than the period
estimated from the pendulum model by about 10{\%} (40{\%}) for the prominence
condensation in the higher (lower) dip.  Higher periods for LAL oscillations
(by up to 20{\%}) than what are estimated from the pendulum model are also
found in the 3D MHD simulation by \citet{Zhou:etal:2018}.
A probable explanation of this discrepancy is that the magnetic field lines
supporting the prominence are
not rigid as assumed in the pendulum model but deform with the oscillations,
which change the curvature radius dynamically \citep{Zhou:etal:2018}.
Figure \ref{fig:2tempfdl_evol_lao} illustrates this deformation by showing a
tracked prominence carrying field line at two time instances, at $t_1$
when the prominence condensation is at the bottom of the dip, and at $t_2$ when
the prominence has moved to the right most position in its oscillation.
We find that with the motion of the prominence to the right, the field line
also slightly swing to the right with the dip location tending to follow
the prominence such that the prominence does not rise as high as it would have
if it had moved strictly along the original field line curve.
This effectively reduces the curvature experienced by the prominence moving
trajectory and thus increasing the oscillation period, which is also found
in the 2D simulations by \citet{Zhang:etal:2019}.
This effect is expected to be significant if the magnetic dips supporting the
prominence are significantly non-force-free because of the prominence weight
\citep{Zhou:etal:2018}, which
is the case for our simulated prominence-carrying flux rope (F18).
The ``pendulum model'' is a first approximation for the LAL oscillations,
and there are other complications, for example the effect of gas pressure gradient as
an additional restoring force \citep{Luna:Diaz:Karpen:2012, Adrover:etal:2020},
although that effect would tend to reduce the oscillation period.

\begin{figure}[htb!]
\centering
\includegraphics[width=0.75\textwidth]{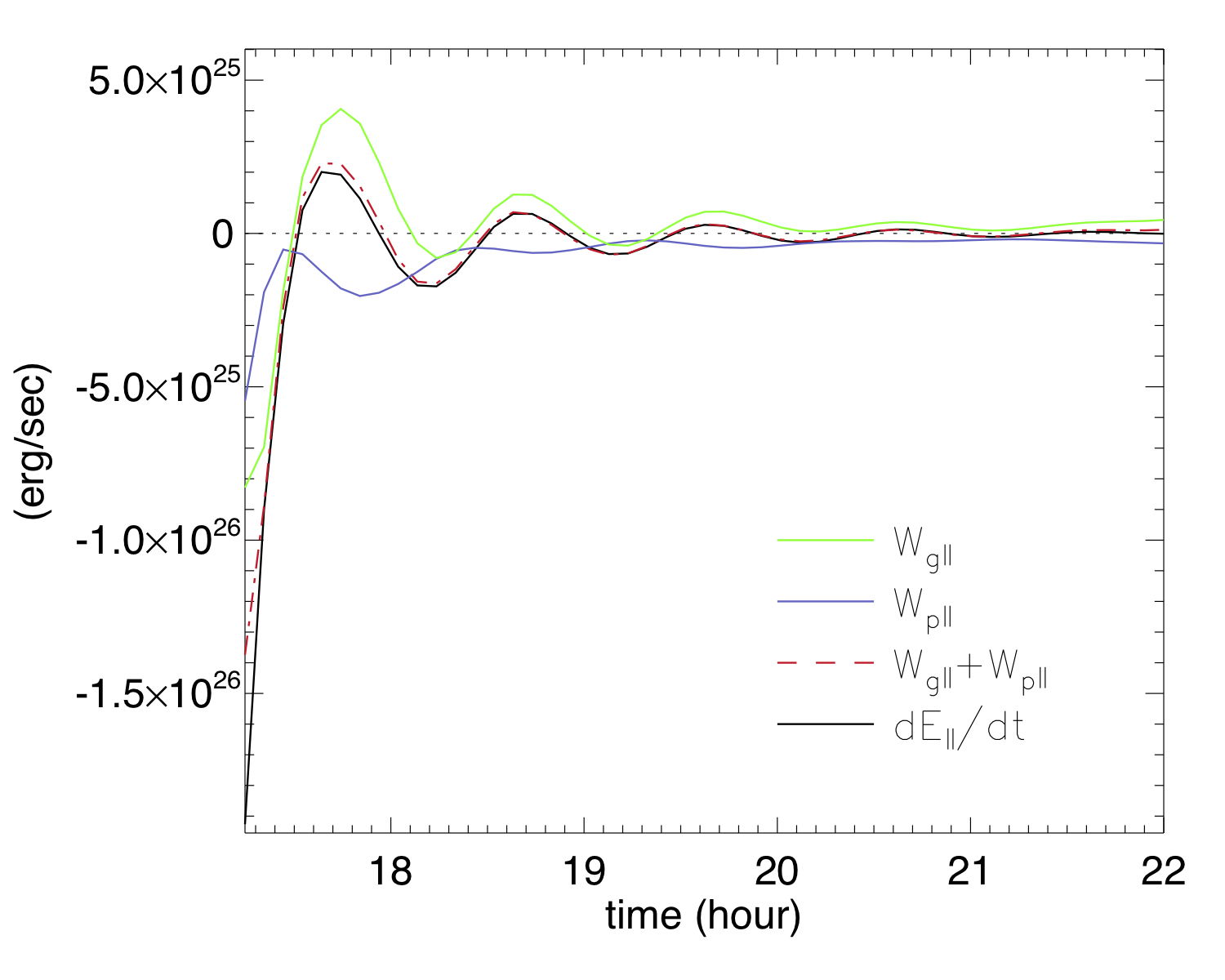}
\caption{The rate of change of the parallel kinetic energy of the prominence
$d E_{\parallel} / dt$ (black curve), and the rate of work done to the prominence
by the field-aligned component of the gravity force $W_{g \parallel}$ (green curve),
the field-aligned component of the pressure gradient force $W_{p \parallel}$ (blue curve)
and the sum of the two $W_{g \parallel} + W_{p \parallel}$ (red dashed line). See text for
expressions for $E_{\parallel}$, $W_{g \parallel}$, and $W_{p \parallel}$.}
\label{fig:energetics_lao}
\end{figure}

Figure \ref{fig:energetics_lao} (black curve) shows the rate of change of
the parallel kinetic energy of the prominence $d E_{\parallel}/dt$, where
\begin{equation}
E_{\parallel} = \int \frac{1}{2} \rho {{\bf v}_s}^2 \, dV_{\rm prom},
\label{eq:E_parallel}
\end{equation}
is the prominence kinetic energy due to the velocity component
${\bf v}_s = ({\bf v} \cdot {\bf B}) {\bf B} / B^2$ parallel
to the magnetic field ${\bf B}$, and the integration above is over the
volume $V_{\rm prom}$ of the cool prominence plasma with $T < 10^5$ K.
In comparison, Figure \ref{fig:energetics_lao} also shows the rate of work done
to the prominence by the field-aligned component of the gravity force (green
curve):
\begin{equation}
W_{g \parallel} = \int {\bf f}_g \cdot {\bf v}_s \, dV_{\rm prom},
\end{equation}
the rate of work done to the prominence by the field-aligned component of the
pressure gradient force (blue curve):
\begin{equation}
W_{p \parallel} = \int {\bf f}_p \cdot {\bf v}_s \, dV_{\rm prom},
\end{equation}
and the sum of the two (red dashed curve). In the above,
${\bf f}_g = - \rho (G M_{\odot}/r^2) {\hat {\bf r}}$
is the gravity force and
and ${\bf f}_p = - \nabla p$ is the pressure gradient force.
It can be seen that, except the very beginning immediately after the initial
velocity is imparted (at about $t=17.15$ hour), the rate of change of the
parallel kinetic energy of the prominence (black curve) is in approximate
agreement with the sum of the rate of work done by the parallel component of the
gravity and pressure gradient forces,
$W_{g \parallel} + W_{p \parallel}$ (red dashed curve), during
the periods of the LAL oscillations.
The gravity part $W_{g \parallel}$ (green curve) shows both positive and negative
rate of work
done to the prominence, roughly in phase with the rate of change of the
parallel kinetic energy $dE_{\parallel} /dt $ (black curve),
indicating that it acts as a restoring force of the
oscillations.  Whereas the pressure gradient part $W_{p \parallel}$ (blue curve) shows
only negative rate of work to the prominence and is significantly out of phase with the
rate of change of the kinetic energy, indicating it is mainly a dissipative
force of the oscillations, and can account for most of the dissipation over
the oscillation periods (except at the very beginning).
At the beginning immediately after the initial velocity is imparted (at about
$t=17.15$ hour), the dissipation of the kinetic energy $dE_{\parallel}/dt$ (black
curve) is significantly greater than the sum of the rate of work by the field-aligned gravity
and pressure gradient $W_{g \parallel} + W_{p \parallel}$ (red dashed curve).  The initial
strong dissipation of the kinetic energy is most likely due to the strong numerical
viscosity in the code produced by the discontinuous initial velocity field imparted
to the prominence.  However the agreement between $dE_{\parallel}/dt$
and $W_{g \parallel} + W_{p \parallel}$
and the fairly close in-phase relation between $dE_{\parallel}/dt$ and $W_{g \parallel}$ during
the LAL oscillations indicate that the field-aligned gravity is the main restoring
force of the LAL oscillations with the field-aligned pressure gradient acting mainly as a drag or
dissipation, consistent with the ``pendulum model'' of \citet{Luna:Karpen:2012}.

\begin{figure}[htb!]
\centering
\includegraphics[width=0.75\textwidth]{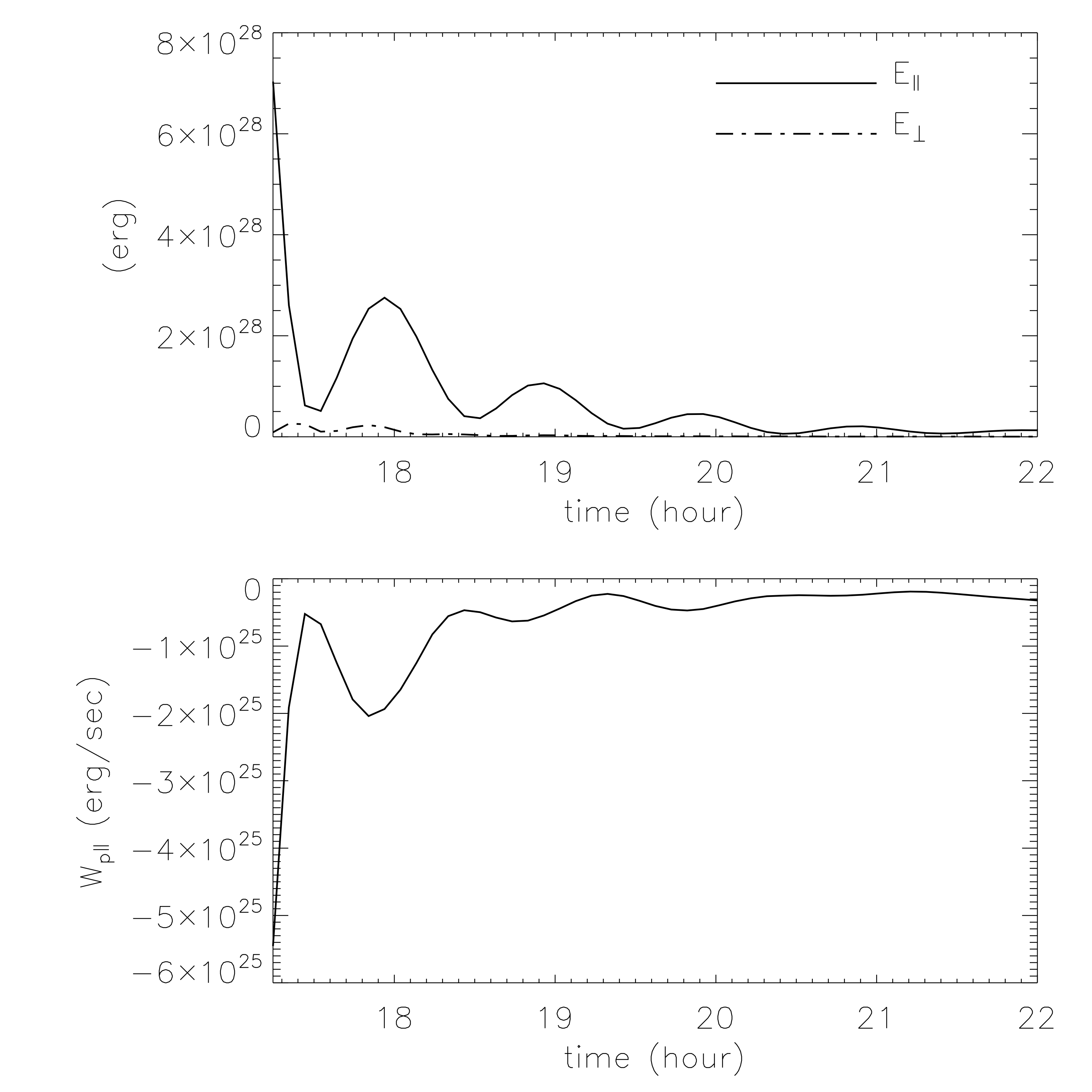}
\caption{(upper panel) The parallel kinetic energy $E_{\parallel}$ (solid line) and
the perpendicular kinetic energy $E_{\perp}$ of the prominence; (lower panel) The rate
of work done by the field-aligned pressure gradient $W_{p \parallel}$.}
\label{fig:workp_eks}
\end{figure}

The damping role of the field-aligned pressure gradient on the LAL oscillations of
the prominence is further illustrated in Figure \ref{fig:workp_eks}, which shows an approximate
anti-phase relation between the parallel kinetic energy $E_{\parallel}$ (in the upper panel)
and the rate of work by the parallel pressure gradient $W_{p \parallel}$ (in the lower panel). 
The consistent negative sign of $W_{p \parallel}$ and its anti-phase relation with $E_{\parallel}$
show that the field-aligned pressure gradient is always acting against the velocity direction,
and its amplitude is well correlated with the magnitude of the velocity, making it behave like
a frictional force.  This result is in agreement with the findings from the simulations by
\citet{Zhang:etal:2019}, who showed that with the inclusion of the non-adiabatic effects,
the pressure gradient becomes a predominantly frictional force instead of a restoring force. 

\begin{figure}[htb!]
\centering
\includegraphics[width=0.75\textwidth]{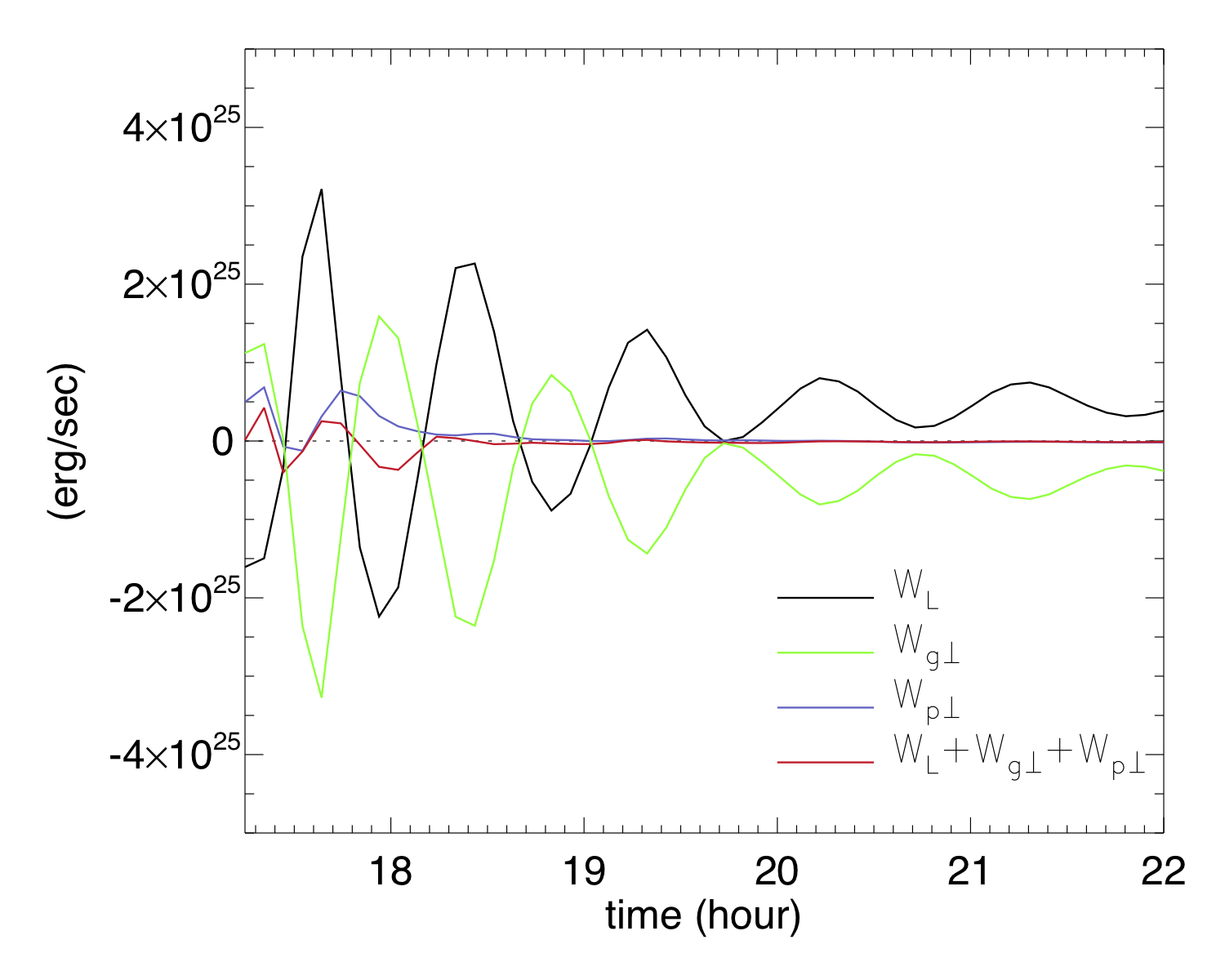}
\caption{The rate of work done to the prominence by the Lorentz force, $W_L$ (black curve),
the perpendicular component of the gravity force $W_{g \perp}$ (green curve), the perpendicular
component of the pressure gradient force $W_{p \perp}$ (blue curve), and their sum (red curve).} 
\label{fig:workperp}
\end{figure}

The upper panel of Figure \ref{fig:workp_eks} also shows that the perpendicular kinetic energy of the
prominence:
\begin{equation}
E_{\perp} = \int \frac{1}{2} \rho {{\bf v}_{\perp}}^2 \, dV_{\rm prom},
\label{eq:ekperp}
\end{equation}
where ${\bf v}_{\perp} = {\bf v} - ({\bf v} \cdot {\bf B}) {\bf B} / B^2 $ is the velocity
component perpendicular to the magnetic field, remains very small compared to the parallel
kinetic energy $E_{\parallel}$.
In other words, there is not a significant excitation of the prominence
transverse oscillations due to the LAL oscillations.
We have also evaluated the rate of work done to the prominence mass by the Lorentz force
${\bf f}_L$ (which is itself perpendicular to the magnetic field line):
\begin{equation}
W_{L} = \int {\bf f}_L \cdot {\bf v} \, dV_{\rm prom},
\end{equation}
and the rate of work done to the prominence by the perpendicular component of the gravity force,
\begin{equation}
W_{g \perp} = \int {\bf f}_g \cdot {\bf v}_{\perp} \, dV_{\rm prom},
\end{equation}
and by the perpendicular component of the pressure gradient force:
\begin{equation}
W_{p \perp} = \int {\bf f}_p \cdot {\bf v}_{\perp} \, dV_{\rm prom},
\end{equation}
and these are shown in Figure \ref{fig:workperp}.
It is found that the rate of work done by the Lorentz force on the prominence (black curve)
is mainly counteracting that by the perpendicular component of the gravity force (green curve).
Overall, the sum of the rate of work done by all of the perpendicular forces
($W_L + W_{g \perp} + W_{p \perp}$) on the prominence, shown as the red curve
in Figure \ref{fig:workperp}, is of a significantly smaller amplitude compared
to the sum of the rate work done by the field-aligned forces (the red dash-dotted curve
in Figure \ref{fig:energetics_lao}), which can account for most of the rate of change
of the kinetic energy of the prominence (the black dash-dotted curve in
Figure \ref{fig:energetics_lao}). Thus the damping of the LAL oscillation kinetic
energy is mainly through the friction-like field-aligned pressure gradient force and also
the numerical viscosity at the beginning.
We note that future higher resolution 3D simulations that significantly reduce the
numerical diffusion are needed to see how well the above result in regard to the
damping of the LAL prominence oscillations in a 3D configuration holds.

\begin{figure}[htb!]
\centering
\includegraphics[width=0.5\textwidth]{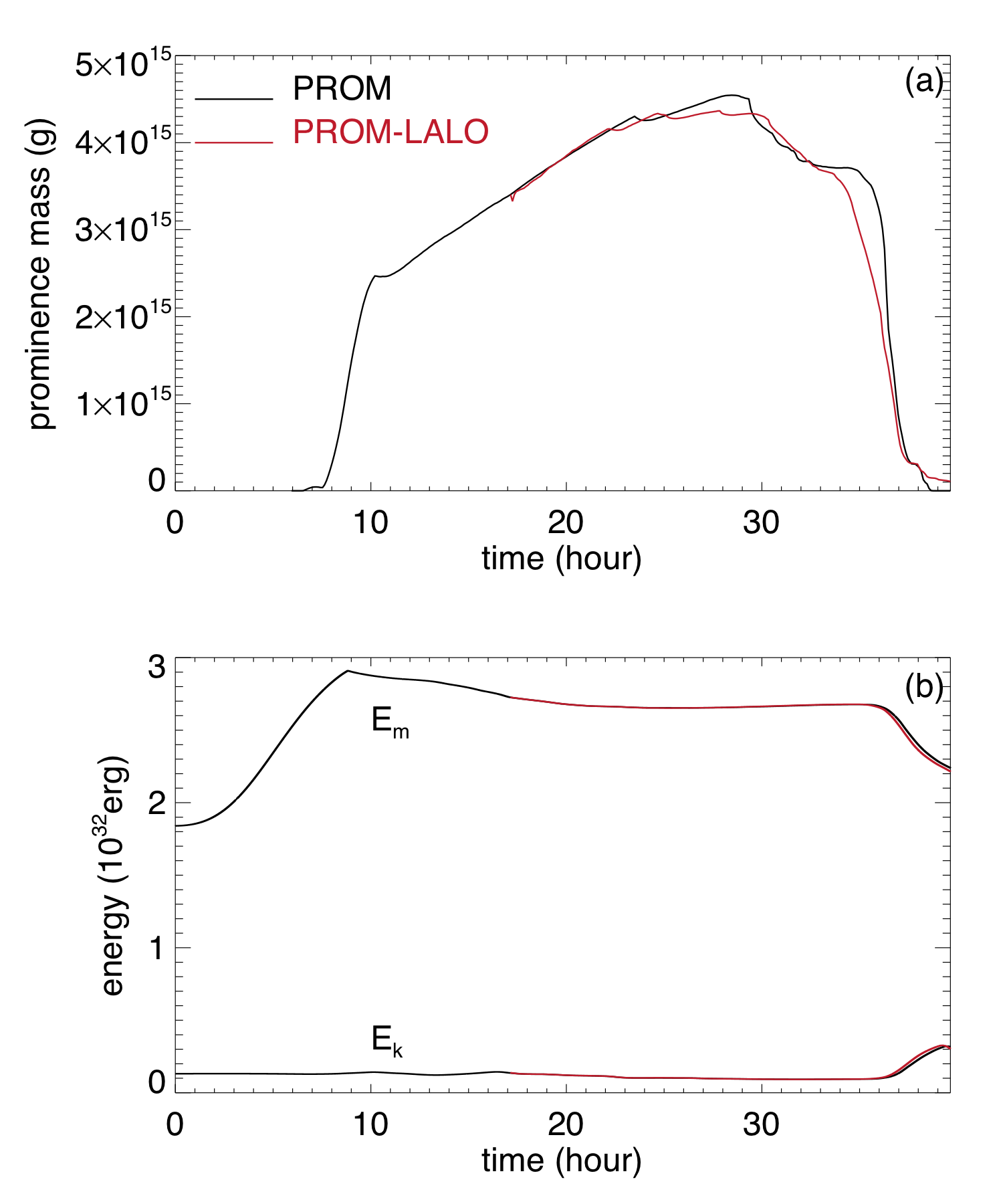}
\caption{(a) The temporal evolution of the cool prominence mass evaluated as the
total mass with temperature below $10^5$ K; (b) The temporal evolution of the
total magnetic energy $E_m$ and total kinetic energy $E_k$.  The results
for the ``PROM-LALO'' case (``PROM'' case) are shown as the red curves
(black curves)}
\label{fig:prommass_evol_lao}
\end{figure}

We find in this simulation that the flux rope is very stable to the
introduction of the LAL oscillations and remain in a stable quasi-static rise after
the oscillations are damped out (Figure \ref{fig:lao_fdl_304} and the associated
movie). The LAL oscillations are followed by episodes of
asymmetric prominence draining towards either footpoints, as the
quasi-static rise makes some of the prominence dips sufficiently shallow.
Figure \ref{fig:prommass_evol_lao} shows the temporal evolution of the
prominence mass and the total magnetic and kinetic energies for the
``PROM-LALO'' case compared to the original ``PROM'' case without
the introduction of the oscillations.
We find that although the temporal and spatial patterns of the
prominence draining episodes are changed
(see the movie for Figure \ref{fig:lao_fdl_304}) compared to
the ``PROM'' case (shown in the movie for Figure \ref{fig:prom_nonprom_fdl_304}),
there is not a significant enhancement 
of the draining of the total prominence mass with the introduction
of the LAL oscillations in the ``PROM-LALO'' case compared to the ''PROM'' case
(Figure \ref{fig:prommass_evol_lao}(a)).
As a result the flux rope rises quasi-statically to the loss-of-equilibrium
height and develops an eruption with a significant release of the magnetic
energy and increase of kinetic energy, at approximately the same time for the
two cases (Figure \ref{fig:prommass_evol_lao}(b)).
We also note that since the prominence LAL oscillations are damped out quickly during
the stable phase before the onset of eruption in our ``PROM-LALO'' simulation,
we do not find a significant increase of the LAL oscillation period with time
as was found in the observational study by \citep{Bi:etal:2014}.  
This is because over the course of the LAL oscillations in our simulation,
the prominence carrying field line dips have not risen significantly
to produce a significant change of the radius of curvature and hence a
significant change of the oscillation period. However, the fact that we
see a larger oscillation period for the higher part of the prominence in the
shallower dips than the lower part in the deeper dips suggests that one may
see the coupling between the oscillation period with the rise of the prominence
if the oscillations are less damped to last through the significant rise,
or if they are introduced at a later phase
when the magnetic field configuration is changing more rapidly as it approaches
eruption. Further 3D simulations are needed to study the coupling between LAL
oscillations and the eruption.

\section{Summary and discussion}
In this paper, we have expanded upon a previous study of the eruption
of prominence-forming coronal flux rope (F18),
and carried out further simulations to investigate the role of prominence
draining and prominence LAL oscillations.  In the previous study of F18, by comparing
simulations with and without the formation of the prominence (the ``PROM''
and ``non-PROM'' simulations in F18), it was shown that prominence weight
can suppress the development of the kink instability of the flux rope and
delays its rise to the loss-of-equilibrium height to develop an eruption.
Further comparison of those two simulations in this paper shows that 
the prominence weight also causes a significant increase of the
loss-of-equilibrium height of the flux rope, consistent with the prediction
from an analytical model by \citet{Jenkins:etal:2019}.
The reason is that the addition of the prominence weight causes the total
confining force of the flux rope to decline more slowly with height,
such that the flux rope needs to rise quasi-statically to a
higher height where the corresponding potential field has a steeper
decline rate (than that is needed in the absence of the prominence weight)
for the loss of equilibrium or torus instability to take place.
A further simulation (the ``PROM-drain'' simulation in this paper), where
we artificially initiate prominence draining during the quasi-static rise
phase of the previous ``PROM'' simulation, shows that a significant
reduction of the total prominence mass compared to the ``PROM'' case,
allows the flux rope to rise more quickly to the loss of equilibrium height
and develop a dynamic eruption significantly earlier (by about 5 hours).
This process of prominence mass-unloading, which causes a speed up of the slow rise
of the flux rope to the loss-of-equilibrium height to then develop a dynamic eruption
is consistent with the evolution seen in the multi-viewpoint observation
of the CME event described by \citet{Seaton:etal:2011}.

We note that the total prominence mass in the ``PROM'' and ``PROM-drain''
simulations reaches a peak value of about $4.5 \times 10^{15}$ g and $3.5 \times
10^{15}$ g respectively during the quasi-static phase of the evolution.
These values are of the same order of magnitude as the high end of the
observed prominence mass range, $\sim 10^{14}$ - $2 \times 10^{15}$ g
\citep[e.g.][]{Parenti:2014}. We have found that the suppression of the kink
instability and the delay of onset of eruption depend on the total prominence
mass.  We have performed a simulation similar to the ``WL-S'' case in F17, where
the constant of proportionality $C$ for the base pressure in equation (18) in F17
used is smaller, such that a much smaller prominence formed in the emerged
flux rope (see Figure 10 in F17) with the peak prominence mass reaching about
$3.6 \times 10^{14}$ g, closer to the lower end of the observed range of prominence
mass.  In this case with a much less massive prominence in the same emerged flux rope
compared to the ``PROM''
and ``PROM-drain'' case, we find that the onset of eruption is only slightly delayed
(by about 0.5 hour) compared to the ``non-PROM'' case, i.e. the stability
and the onset of eruption is not significantly altered by the prominence mass.
This result suggests that even though locally the prominence carrying field is
significantly non-force-free, the total prominence mass
formed also needs to be sufficiently large for it to significantly affect the
stability and the onset of eruption of the flux rope.  More extended 
simulations exploring different flux rope and confining field configurations
are needed to study this necessary prominence mass given the field strength
of the flux rope.

We have also carried out a simulation (the ``PROM-LALO'' simulation in this
paper) where we excite LAL oscillations by adding an initial velocity (parallel to the
magnetic field) of 100 km/s to the prominence plasma (with temperature $T< 10^5$ K) 
at a time instant during the quasi-static phase of the ``PROM'' simulation,
and study the subsequent evolution in comparison to the original ``PROM''
simulation.  We find prominence LAL oscillations develop with an oscillation period of
roughly 2 hours. The prominence shows differential motions with the upper part
of the prominence in shallower dips showing a slightly longer oscillation period 
than that in the lower deeper dips.  The oscillation periods are found to be
longer (by about 10\% to 40\%) than those estimated based on the ``pendulum model''
\citep{Luna:Karpen:2012} using the mean radius of curvature of the field line dips.
The 3D simulation of LAL oscillations by \citet{Zhou:etal:2018} have also found
longer periods than expected by the pendulum model.
A probable cause for this is that the magnetic field lines supporting the prominence are
not rigid as assumed in the pendulum model but deform with the oscillations.
The deformation tends to drag the dip location along with the prominence motion,
effectively reducing the curvature experienced by the trajectory of the the moving
prominence condensation and hence increasing the oscillation period.
This effect needs to be taken into account when one uses the observed LAL oscillation
periods to estimate the curvature radius and the magnetic field strength for the field
line dips supporting the prominence based on the pendulum model
\citep{Luna:Karpen:2012,Bi:etal:2014}.
It would tend to give an over estimate of the magnetic field strength.
It is found that the rate of change of the parallel kinetic energy
of the prominence during the LAL oscillations can be well described by the sum of
the work done by the parallel components of the gravity force and the pressure
gradient force, with the gravity force acts as the main restoring force and
the pressure gradient mainly as a dissipation.  The oscillations are found
to be strongly damped and die out after about 2-3 oscillation periods.
The flux rope is stable with the introduction of the LAL oscillations, and remains
in the quasi-static rise after the oscillations are damped out.
It develops subsequent episodes of prominence draining with spatial and temporal
patterns different from that in the original ``PROM'' case, but the overall draining
of the total prominence mass is not significantly enhanced, and it develops the final
eruption at approximately the same time as that in the ``PROM'' case.  Thus, even though
both the LAL oscillations and draining develop prior to the eruption, as is found
in the observation of \citet{Bi:etal:2014}, the oscillations do not significantly
alter the overall rate of prominence draining and the initiation of eruption
in this case.

\acknowledgments
The author would like to thank the anonymous referee for helpful comments that
improved the paper.
This material is based upon work supported by the National Center for Atmospheric
Research, which is a major facility sponsored by the National Science Foundation
under Cooperative Agreement No. 1852977. 
This work is also supported in part by the NASA LWS grant 80NSSC19K0070 and by
the Air Force Office of Scientific Research grant FA9550-15-1-0030 to NCAR.
We would like to acknowledge high-performance computing support from Cheyenne
(doi:10.5065/D6RX99HX) provided by NCAR's Computational and Information Systems
Laboratory, sponsored by the National Science Foundation.
Y.F. would like to thank the International Space Science Institute (ISSI)
team on ``Large-Amplitude Oscillations as a Probe of Quiescent and Erupting Solar
Prominences'' led by Manuel Luna for support to attend the team meetings and for
helpful discussions that initiated this work.

\end{document}